\documentclass[prb,preprint,showpacs,preprintnumbers,amsmath,amssymb]{revtex4}
\usepackage{graphicx}
\usepackage{amssymb}
\usepackage{mathrsfs}
\usepackage{dcolumn}
\usepackage{bm}

\begin{document}
\title{Spin Frank-Condon Effect}

\author{Mahrous R. Ahmed\footnote{Permanent address:  Department of Physics, Faculty of Science, South Valley University, 82524 Sohag, Egypt.}}
\email[Corresponding author: ]{php02mra@shef.ac.uk,  mahrous_r_ahmed@yahoo.com}
\author{G. A. Gehring}

\affiliation{Department of Physics and Astronomy, University of Sheffield, Hicks Building, Hounsfield Road, Sheffield, S3 7RH, UK.}

\date{\today}

\begin{abstract}
The optical properties of a small magnetic cluster are studied in a magnetic version of Frank-Condon principle. This simple model is considered to show new basic physics and could be adopted to treat real problems. The energies and wavefunctions of the cluster are calculated for different spin configurations to evaluate the energies and the strengths of the allowed transitions from the relaxed excited states. The optical de-excitation energies for the likely scenarios are obtained in terms of the exchange parameters of the model. 
\end{abstract}


\maketitle
\parindent1cm

\section{\label{2}Introduction}
\begin{figure}
\begin{center}
\includegraphics[scale=0.5] {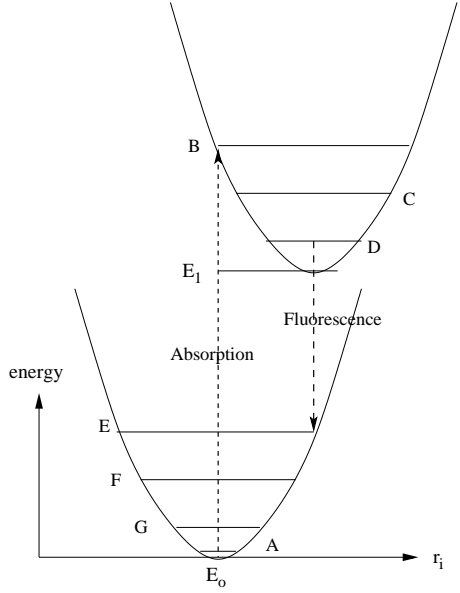}
\end{center}
\caption{\label{fc}Energy diagram of an electronic transition with phonon coupling along the configurational coordinate $r_i$, a normal mode of the lattice. The upwards arrows represent absorption without phonons. The downwards arrows represent the symmetric process in emission.}
\end{figure}

In small molecules, excited electronic states may have altered atomic coordinations, which leads to Frank-Condon multi-phonon sidebands in electronic spectra\cite{Herzberg50}. In solids, electronic excited states are often delocalised, eliminating such effects\cite{Rashba82}. However, if excited states are self-localised\cite{Rashba, Sog93} then Frank-Condon effects should reappear in the form of shifting and Gaussian broadening of the pure electronic transition\cite{Cho75}. The electronically excited cluster may relax to its vibrational ground state after excitation so that the fluorescence energy is lower than the energy required for absorption. This is sketched in figure \ref{fc}. In this paper we explore the possibility of a spin Frank-Condon effect and describe what would be observed in this case.

We consider a spin cluster such that after excitation the lowest energy spin configuration differs from that in the ground state and so a spin relaxation may occur.  We calculate the energies and wavefunctions of the cluster for different spin configurations.  This is used to evaluate the energies and the strengths of the allowed transitions from the ground state and from the relaxed excited states.

In the manganites it is known that the exchange interaction is antiferromagnetic between two Mn$^{3+}$ ions but becomes ferromagnetic\cite{Wollan55} if one of the atoms is ionised to Mn$^{4+}$.  This gives the possibility for the exchange to depend on the state of excitation of the Mn ions.  Another way in which a scenario similar to the one that we describe can occur is if the spin that we consider is actually a pseudo-spin\cite{Ahmed06} corresponding to the orbital order in manganites\cite{Brink02}. In this case ionising the Mn$^{3+}$ ion will eliminate the orbital order on that site and could also lead to a realignment of the orbital  moments on the neighbouring ions\cite{Allen99}. In this paper we consider a toy model that shows this basic physics and can be adapted to treat the real problems mentioned above.

We summarise the procedure for calculating Frank-Condon spectra and then describe how the spin calculation proceeds along the same path. First we calculate the vibrational energies and wavefunction of the cluster in the electronic ground state and then repeat the calculation for the excited cluster. This enable us to see when the excited state is off-set relative to the ground state as shown in figure \ref{fc}. In this case the excited cluster may make radiationless transitions to the vibrational ground state of the electrically excited cluster. The Frank-Condon principle states that the transition is a vertical line as shown in the figure. More accurately, the intensity of an electronic transition that is accompanied by changes in the vibrational states is proportional to the square of overlap of the two vibrational wave functions.  

In the spin case we use a model Hamiltonian to calculate the energies and wavefunctions of the spin cluster in the (electronic) ground state and when an electron is excited. We assume that the excited electron is delocalised so that the magnetic cluster is left with a vacancy. This enables us to identify the parameter ranges for which reorientation may occur in the excited state. The electronic transition will occur without spin flip of the neighbouring ions which is the spin analog to the Frank-Condon approximation. The intensity of each transition is then found from the overlap of the two spin states.

In Section II we define the spin cluster model and calculate the energies and the eigenstates for the electronic ground state and when an electron is excited.  The excitation spectra are calculated in section III and the paper concludes in Section IV.

\section{\label{method}Methodology}
\subsection{ The effective Hamiltonian }
\begin{figure}
\begin{center}
\includegraphics[scale=0.6] {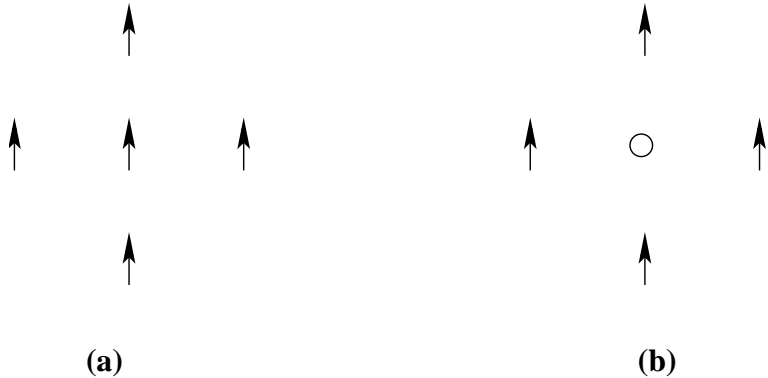} 
\end{center}
\caption{\label{gs}(a) The ground state of ferromagnetic spin cluster with total spin $S_T=5/2$, (b) the cluster in a state with total spin $S_T=2$ just after the central spin is removed by optical excitation.}
\end{figure}

We consider a cluster of five spins as shown in figure \ref{gs}a. In the ground state it has the highest spin total $S_T=5/2$. We assume that after optical excitation an electron for the central site has been excited and has left the cluster as shown in figure 2b which is a state of ($S_T=2$). It will be seen that this toy model is very rich and allows us to explore this effect in detail. 

The cluster de-excites when a band electron makes a transition back on to the central site.  We shall consider both possibilities namely that the spin of the electron that rejoins the cluster is parallel or antiparallel to the spin of the whole cluster.

We assume that there is a Heisenberg ferromagnetic interaction, $J'$, between the central spin and the four neighbours and an antiferromagnetic Heisenberg interaction, $J$, between the neighbours and that each of the four neighbours experiences a mean field interaction, $H_{mf}$, with the rest of the lattice.
 
The effective Hamiltonian describing the low-energy electronic spin states is given by,
\begin{equation}
{\mathcal{H}} = H_0+H_1+H_2,
\end{equation}
where $H_o$ represents the crystal mean field interaction. Then,
\begin{equation}
H_o=-H_{mf}(S_{1}^{z}+S_{2}^{z}+S_{3}^{z}+S_{4}^{z}),
\end{equation}
\begin{eqnarray}
H_1&=&-J'{\bf S_o}.({\bf S_1}+{\bf S_2}+{\bf S_3}+{\bf S_4})
\end{eqnarray}
and
\begin{equation}
H_2=J({\bf S_1}.{\bf S_2}+{\bf S_2}.{\bf S_3}+{\bf S_3}.{\bf S_4}+{\bf S_4}.{\bf S_1}).  
\end{equation}
Where ${\bf S_o}$ is the spin operator of the spin in the centre of the cluster and ${\bf S_1}, {\bf S_2}, {\bf S_3}$ and ${\bf S_4}$ are the spin operators of the nearest neighbours of the central spin. 

After the central spin, $S_o$, is removed from the cluster as shown in figure \ref{gs}b the term $H_1$ does not contribute. The effective Hamiltonian representing the excited cluster under the crystal mean field can be given as follows,
\begin{eqnarray}
{\mathcal{H}}&=&-H_{mf}(S_{1}^{z}+S_{2}^{z}+S_{3}^{z}+S_{4}^{z})+
J(S_{1}^{z}S_{2}^{z}+S_{2}^{z}S_{3}^{z}+S_{3}^{z}S_{4}^{z}+S_{4}^{z}S_{1}^{z}) \nonumber \\     &&+(J/2)(S_{1}^{+}S_{2}^{-}+S_{1}^{-}S_{2}^{+}+S_{2}^{+}S_{3}^{-}+S_{2}^{-}S_{3}^{+} \nonumber \\ &&+S_{3}^{+}S_{4}^{-}+S_{3}^{-}S_{4}^{+}+S_{4}^{+}S_{1}^{-}+S_{4}^{-}S_{1}^{+}),
\end{eqnarray}
where $S_{1}^{z}, S_{2}^{z}, S_{3}^{z}$, and $S_{4}^{z}$ are the spin-up and the spin-down operators. The parameters are chosen so that the ground state of the five-spin cluster is ferromagnetic. The energy of the state with total spin $S_T=5/2$ is $E_{5/2}=-4H_{mf}-4J'+4J$. The energies and eigenstates of the states $S_T=3/2$, $S_T=1/2$ and  $S_T=-1/2$ are given in the table \ref{3/2}. 

\begin{figure}[h!]
\begin{center}
\includegraphics[scale=0.4] {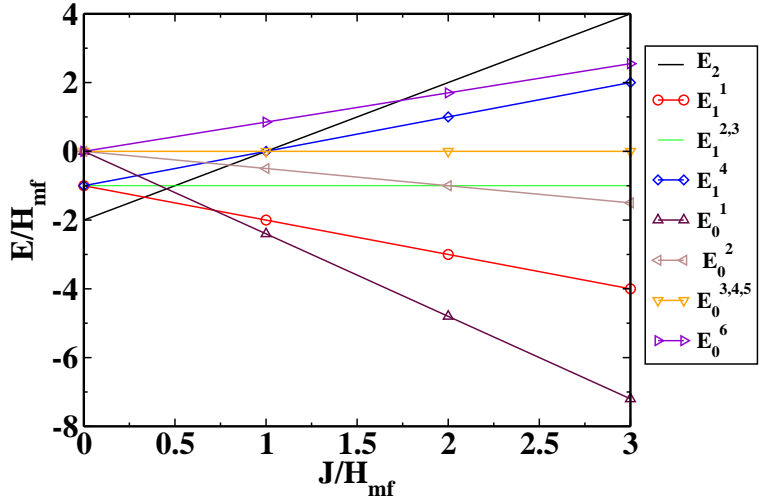}
\end{center}
\caption{\label{E_J}(a) A diagram shows the eigen energy $E/H_{mf}$ versus the AFM exchange energy $J/H_{mf}$. $E_2$ is ground state at $0<J/H_{mf}<0.33$ range, $E_0^1$ is the ground state at the range $0.33<J/H_{mf}<0.72$ and finally $E_1^1$ is the ground state above this range $0.72<J/H_{mf}$.}
\end{figure}

Similarly calculations are done for the four-spin cluster. Immediately after excitation it will be in the state $S_T=2$, as shown in figure \ref{gs}b. The energies of the four-spin cluster are shown in figure \ref{E_J} as a function of $J/H_{mf}$. There is one state of $S_T=2$, four states of $S_T=1$ (two are degenerate) and six states of $S_T=0$ (three are degenerate). Since the cluster relaxes to the lowest state we are most interested in the energy and configuration of the lowest state. It is seen that the spin of the ground state of this four-spin cluster changes as a function of $J/H_{mf}$. For small values of $J$ such that $0<J/H_{mf}<0.33$ the spin of the ground state remains as $S_T=2$ and then there is an intermediate region $0.33<J/H_{mf}<0.72$ where the ground state has $S_T=1$ (state $E_1^1$) and finally for large values of $J/H_{mf}$ it becomes $S_T=0$ (state $E_0^1$).

\begin{table}[h!]
\caption{The eigen energies of the states with total spin $S_T=3/2$, $S_T=1/2$ and $S_T=-1/2$.}
\label{3/2}
\begin{tabular*}{\textwidth}{@{\extracolsep{\fill}}ll}
\hline
$S_T$           &Eigen energy                                   \\
\hline
$S_T=3/2$    & $E_{3/2}^{1,2}=\frac{1}{4}(-6H_{mf}+J'$          \\
             &$-\sqrt{16J^2+24JJ'+2JJ'^2})$                     \\
             &$E_{3/2}^{3,4}=\frac{1}{2}(-3H_{mf}-4J')$         \\
             &$E_{3/2}^5=\frac{1}{2}(-3H_{mf}+2J-J')$           \\
             
\hline
$S_T=1/2$    &$E_{1/2}^{1,2,3}=0$                \\              
             &$E_{1/2}^4=0$                  \\
             &$E_{1/2}^{5,6}=\frac{1}{2}(-2H_{mf}+2J'-\sqrt{2}\sqrt{2H_{mf}^2-4H_{mf}J'+3J'^3}$                   \\
             &$E_{1/2}^{7,8}=\frac{1}{2}(-2H_{mf}+2J'+\sqrt{2}\sqrt{2H_{mf}^2-4H_{mf}J'+3J'^3})$                     \\
             &$E_{1/2}^{9,10}=\frac{1}{2}(-2H_{mf}-3J+2J'-\sqrt{4H_{mf}^2-20H_{mf}J'+25J^2-8H_{mf}J'+20JJ'+6J'^2)}$  \\
\hline
$S_T=-1/2$   &$E_{1/2}^{1,2,3}=0$            \\
             &$E_{1/2}^4=0$                  \\
             &$E_{1/2}^{5,6}=\frac{1}{2}(2H_{mf}+2J'-\sqrt{2}\sqrt{2H_{mf}^2+4H_{mf}J'+3J'^3}$                   \\
             &$E_{1/2}^{7,8}=\frac{1}{2}(2H_{mf}+2J'+\sqrt{2}\sqrt{2H_{mf}^2+4H_{mf}J'+3J'^3})$                     \\
             &$E_{1/2}^{9,10}=\frac{1}{2}(2H_{mf}-3J+2J'-\sqrt{4H_{mf}^2+20H_{mf}J'+25J^2+8H_{mf}J'+20JJ'+6J'^2)}$  \\
\hline
\hline
\end{tabular*}
\end{table}
We have three different scenarios, I, II and III corresponding to the de-excitation from a cluster of total spin $S_T=2,1$ and 0 respectively. As a band electron combines with the four spin cluster it will produce a state that is the direct product of the state for the four spin cluster and the spin of the band electron. It is this state whose overlap with the ground state wavefunctions must be evaluated in order to obtain the strength of the transition.
 
In scenario I where $0<J/H_{mf}<0.33$ the four-spin cluster remains in the state $S_T=2$. The cluster can de-excite by absorbing an electron of either spin; in the first case it will go back to the ground state of the five spin cluster with $S_T=5/2$, and in the second to an excited state of the ground cluster with $S_T=3/2$. This is shown in figure \ref{levels}a.

In scenario II which is valid for $0.33<J/H_{mf}<0.72$ the four spin cluster first relaxes to the lowest $S_T=1$ state, $E_1^1$, before combining with a band electron of either spin to de-excite to a state with $S_T=3/2$ or $S_T=1/2$. This is shown in figure \ref{levels}b. Finally in the case where $J/H_{mf}>0.72$ the four spin cluster first relaxes to the lowest $S_T=0$ state, $E_0^1$, before combining with a band electron of either spin to de-excite to a state with $S_T=1/2$ or $S_T=-1/2$. This is shown in figure \ref{levels}c.

In the following section we calculate the energies and probabilities corresponding to each of these processes. We note that for the four spin cluster we only need to calculate the wavefunctions for the lowest eigenstate for each value of the total spin. We also need the energies of the five spin cluster for those states that will be reached by de-excitation from the four spin cluster.

\section{\label{results}Results and Discussion}
\subsection{Spin de-excitation}
\begin{figure}[h!]
\begin{center}
\includegraphics[scale=0.8]{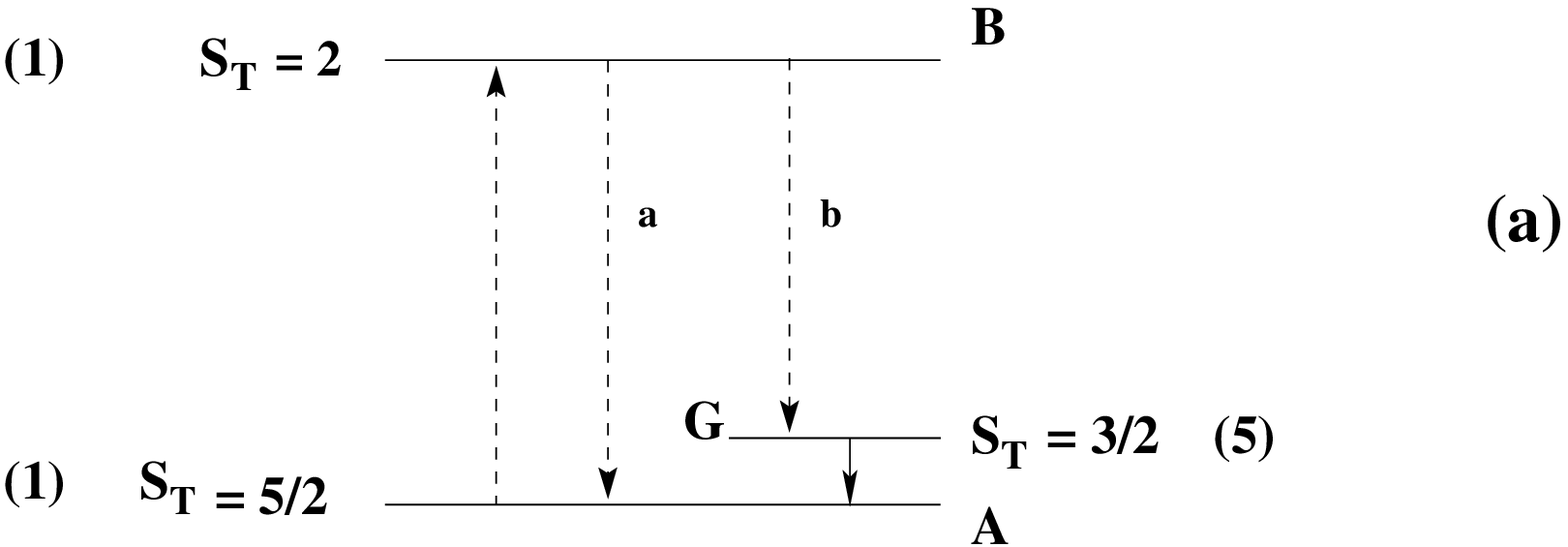}\\
\vspace{1cm}\includegraphics[scale=0.8]{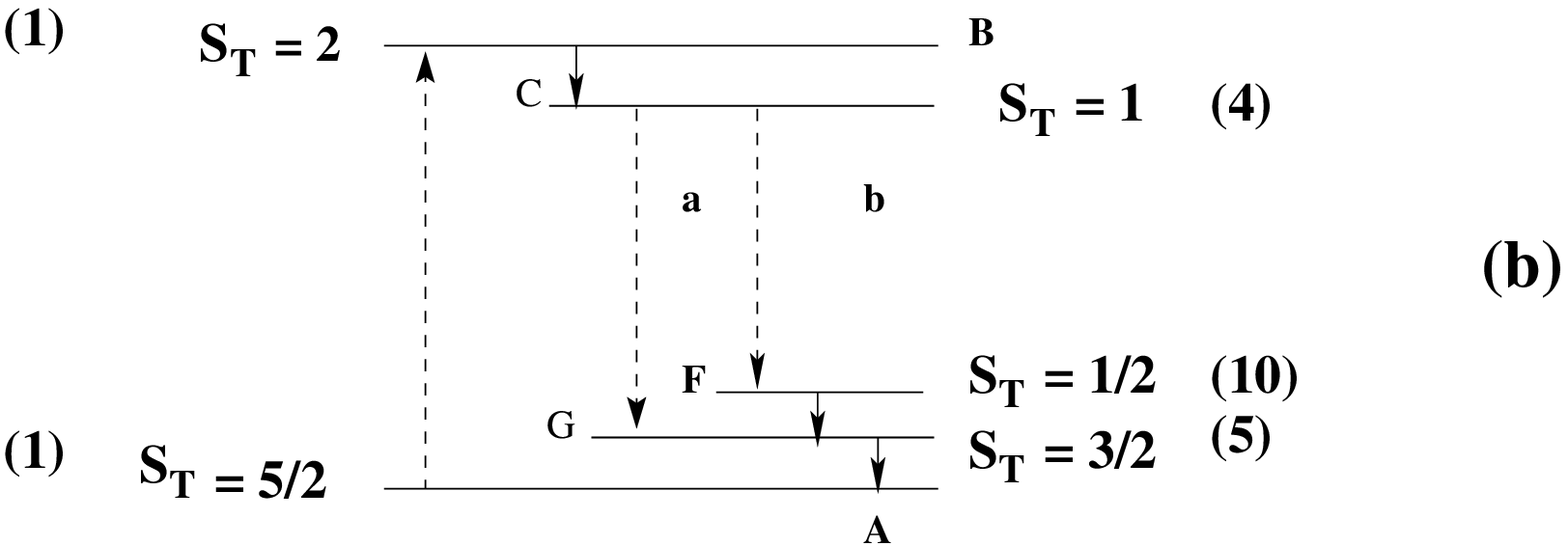}\\
\vspace{1cm}\includegraphics[scale=0.8]{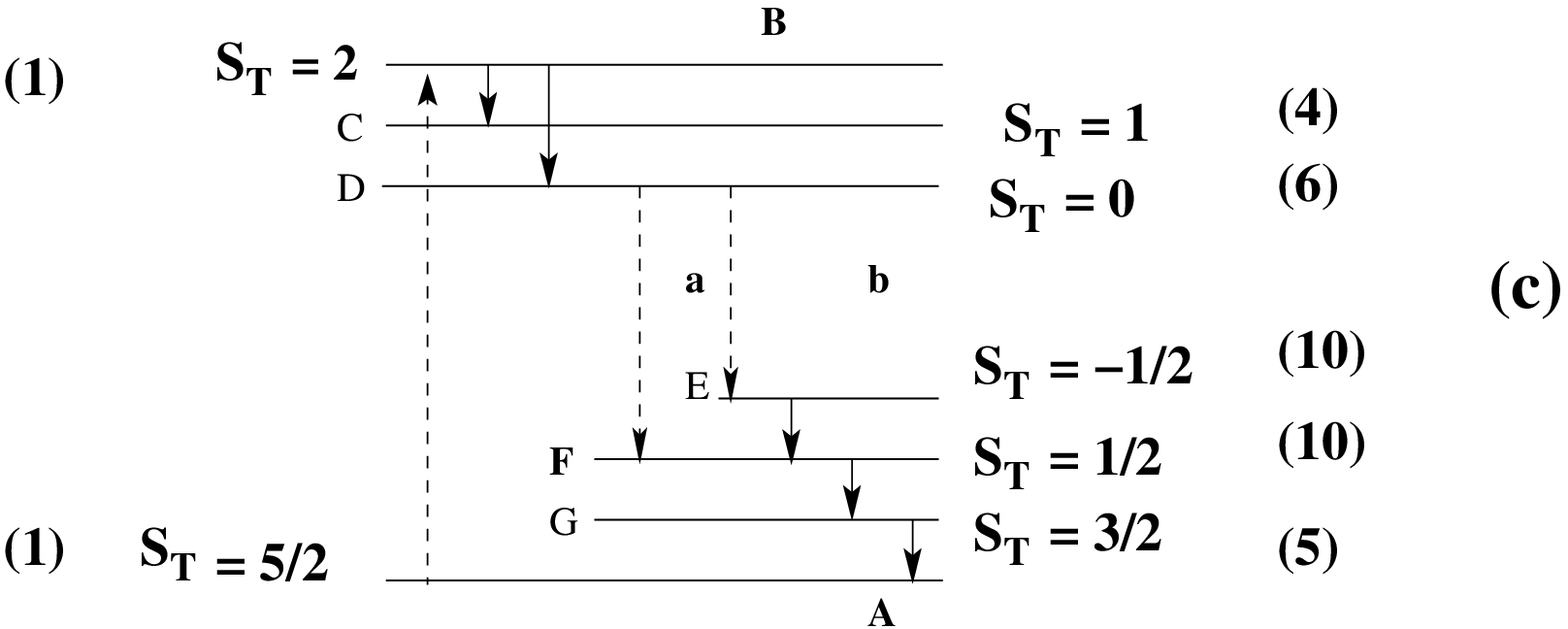}\\
\end{center}
\caption{\label{levels}(a) A diagram describes the relaxation of the cluster according to scenario I, (b) scenario II, (c) scenario III. The number of the levels multiplicity are in brackets.}
\end{figure}

{\bf Scenario I}(a): There are two possibilities for this scenario. First of all, if the band electron combines with the cluster as spin-up, as seen in figure \ref{gs}a, the cluster loses its energy as luminescence and de-excites directly from the energy level-B, $E_2=-4H_{mf}+4J$, with $S_T=2$ to the energy level-A with $S_T=5/2$ as shown in figure \ref{levels}a. Namely, the cluster does not experience any relaxation by losing phonons in this case before or after it optically de-excites which is represented in figure \ref{levels}a by dashed downarrow-a. The eigen energy for these energy levels and their wavefunction are listed in table \ref{210}.

(b): The cluster, in this scenario, starts with the same level $S_T=2$ but the band electron combines with the cluster as spin-down to its central site. The cluster de-excites optically not to the ground state but to an excited state at the energy level-G with $S_T=3/2$. This energy level has five states as shown in table \ref{3/2}. We calculate the overlap of states, $\psi_i (i=1,$...,5), of the five-spin cluster with total spin $S_T=3/2$ with the state formed by the direct product of $\vert\phi_2\rangle$ and an electron of spin down on the central state,$\vert\downarrow\phi_2\rangle$ and the overlap probability, $P_i=\langle\psi_i\vert\downarrow\phi_2\rangle$.

\begin{table}[h!]
\caption{The eigen energies of the states with $S_T=2$, $S_T=1$ and $S_T=0$ and the wavefunctions for the lowest energy state.}
\label{210}
\begin{tabular*}{\textwidth}{@{\extracolsep{\fill}}lll}
\hline
$S_T$           &Eigen energy                      & Wave function                             \\
\hline
$S_T=2$       &$E_2=-4H_{mf}+4J$       &$\psi_2=|\uparrow\uparrow\uparrow\uparrow\rangle$       \\
\hline
$S_T=1$       &$E_1^1=-H_{mf}-J$  &$\psi_1^1=\frac{1}{2}[|\uparrow\uparrow\uparrow\downarrow\rangle+|\uparrow\uparrow\downarrow\uparrow\rangle+|\uparrow\uparrow\downarrow\uparrow\rangle+|\downarrow\uparrow\uparrow\uparrow\rangle]$     \\
             &$E_1^2=-H_{mf}+J$              &             \\              
             &$E_1^3=E_1^4=-H_{mf}$          &        \\
              
\hline
$S_T=0$      &$E_0^1=-2.4J$           &$\psi_0^1=\frac{1}{\sqrt{6}}[|\uparrow\downarrow\uparrow\downarrow\rangle+|\downarrow\uparrow \downarrow\uparrow\rangle+|\uparrow\uparrow\downarrow\downarrow\rangle+|\downarrow\downarrow\uparrow\uparrow\rangle+|\uparrow\downarrow\downarrow\uparrow\rangle+|\downarrow\uparrow\uparrow\downarrow\rangle]$\\
             &$E_0^2=-J/2$                 &           \\
             &$E_0^3=E_0^4=E_0^5=0$       &          \\
             &$E_0^6=0.85J$               &           \\

\hline
\hline
\end{tabular*}
\end{table}
It is found that the cluster de-excites from the energy level-B to the state with energy $E_{3/2}^1=E_{3/2}^2$ with unit probability. The probability of de-excitation of the rest of the states in the spin cluster corresponding to the total spin $S_T=3/2$ are zero.

{\bf Scenario II}: The cluster experiences relaxation by losing thermal energy before it de-excites optically. Namely, before the band electron combines with the cluster, the cluster relaxes thermally from the energy level-B with  $S_T=2$ to the energy level-C with $S_T=1$, as shown in figure \ref{levels}b. The new energy level has four states  which their eigen energy are listed in table \ref{210}. We assume that the cluster relaxes to its lowest energy state $E_1^1$. 

(a) In this scenario the band electron combines with the cluster as spin-up. The cluster de-excites optically with the starting state, $\vert\uparrow\psi_1^1\rangle$ which is direct product of the multiplet of energy level-G with $S_T=3/2$, see figure \ref{levels}b. 

Table \ref{3/2} shows the values of the eigen energy of level-G with $S_T=3/2$ with the starting state, $\vert\uparrow\psi_1^1\rangle$. It is found that the cluster de-excites from the energy level-B to the degenerate level with energy $E_{3/2}=E_{3/2}^1=E_{3/2}^2$ with unit probability. Naturally we fined there is zero probability for de-excitation into any other of five states of five spin cluster with total spin $S_T=3/2$. Finally, the cluster relaxes thermally from the energy level-G with $S_T=3/2$ to the ground state energy level-A with $S_T=5/2$

(b) In this case the cluster absorbs a down spin electron. The cluster de-excites optically from the energy level-C with the starting state, $\vert\downarrow\psi_1^1\rangle$, to an energy level-F higher than the energy level-G. This new level (level-F) has total spin $S_T=1/2$ and has ten states. 

The eigen energies of these ten states of the energy level-F with total spin $S_T=1/2$ have been calculated and shown in table \ref{3/2}. These states from state $E_{1/2}^4$  to $E_{1/2}^10$ have zero overlapping probability with the starting state while the states $E_{1/2}^1$, $E_{1/2}^2$ and $E_{1/2}^{3}$ are overlapping with the starting state by unit probability. Namely, the cluster de-excites optically from the energy level-C to any state of the states from $E_{1/2}^1$ to $E_{1/2}^3$ of the energy level-F, this is represented by the dashed downarrow-b in figure \ref{levels}b. The cluster relaxes thermally to the ground state energy level-A with $S_T=5/2$ through the energy level-G with $S_T=3/2$.

{\bf Scenario III}:(a) After the cluster relaxes from the energy level-B, by losing thermal energy, to level-C with $S_T=1$ it relaxes again to an energy level called D with $S_T=0$. The energy level-D has six states according to the effective Hamiltonian. Table \ref{210} shows the eigen energy for these states. If the band electron combines with the cluster as spin-up, the cluster de-excites optically from this new energy level-D with $S_T=0$ with the starting state, $\vert\uparrow\psi_0^1\rangle$ to the energy level-F with $S_T=1/2$ . 

The energy level-F has ten states.  We have calculated the eigen energies of these ten states with the starting state, $\vert\uparrow\psi_0^1\rangle$ and shown the results in table \ref{3/2}. Our calculations are showing that the states from $E_{1/2}^4$ to $E_{1/2}^10$ have zero overlapping with the starting state. The states $E_{1/2}^1$, $E_{1/2}^2$ and $E_{1/2}^3$ have the maximum probability of overlapping with the starting state. Namely, the cluster de-excites optically from the energy level-D  with $S_T=0$ to the states from 1 to 3 of the energy level-F with $S_T=1/2$, see the dashed downarrow-a in figure \ref{levels}c. Finally the cluster relaxes thermally from the energy level-G to the ground state energy level-A with $S_T=5/2$ through the energy level-G. 

(b): Now, if the band electron combines with the cluster back as spin-down the cluster relaxes optically from the energy level-D with $S_T=0$ to a new energy level called E with $S_T=-1/2$ with new starting state, $\vert\downarrow\psi_0^1\rangle$, as seen in figure \ref{levels}c. It is obtained from our calculations that when the band electron combines with the cluster as spin down during the cluster in the energy level-D with $S_T=0$ the cluster de-excites optically with the starting state, $\vert\downarrow\psi_0^1\rangle$, to the states from $E_{-1/2}^1$ to $E_{-1/2}^3$ of the ten states of the energy level-E with $S_T=-1/2$ where the probability of overlapping of these states with the starting state is the optimum. This is represented by a dashed downarrow-b in figure \ref{levels}c. The states from $E_{1/2}^4$ to $E_{1/2}^{10}$ have zero probability to overlap with the starting state. Finally the cluster relaxes thermally from level-E through levels-F and G to get the ground state energy level-A with $S_T=0$, as seen in figure \ref{levels}c.

\subsection{\label{E_l}The luminescence energy calculation}
We calculate the luminescence energy for each scenario as follows.\\
Scenario I: (a) If the cluster de-excites from energy level-B directly to the ground state energy level-A when the band electron combines with the cluster as spin-up. The luminescence energy $E_l$ for this case is 
\begin{eqnarray}
E_l=E_o&=&E_2-E_{5/2}          \nonumber          \\
       &=&(-4H_{mf}+4J)-(-4H_{mf}-4J'+4J)=4J'.  
\end{eqnarray}
(b) But if the band electron combines with the cluster as spin-down the cluster de-excites first optically from level-B to the states of energy level-G overlapping with $\vert\downarrow\psi_2\rangle$ then it relaxes losing phonons to energy level-A where this relaxation, here, has energy difference $\Delta_1=E_{3/2}^{1,2}-E_{5/2}$. Then the optical energy in this case is, 
\begin{eqnarray}
E_l&=&E_o-\Delta_1              \nonumber            \\
&=&-\frac{19}{2}H_{mf}+8J+\frac{17}{4}J'+\frac{1}{4}\sqrt{16J^2+24JJ'+2JJ'}.
\end{eqnarray}
Scenario-II: The cluster relaxes losing phonons to the level-C with $S_T=1$. The energy difference, here, is $\delta_1=E_2-E_1^1=5H_{mf}-5.5J$, where $E_1^1$ is the lowest eigenstate in level-C to which the cluster relaxes from level-B. The cluster de-excites after that optically to level-G if the band electron combines with the cluster as spin-up and to level-F if the band electron combines with the cluster as spin-down. As we know the energy difference between level-A and level-C is $\delta_1$ and between the level-F and level-A is $\Delta_2=E_{1/2}^{1,2,3}-E_{5/2}$. Then, in the first case, 
\begin{eqnarray}
E_l&=&E_o-\delta_1-\Delta_1               \nonumber    \\
&=&-\frac{29}{2}H_{mf}+\frac{51}{5}J+\frac{17}{4}J'+\frac{1}{4}\sqrt{16J^2+24JJ'+2JJ'}, 
\end{eqnarray}
but in the second case, 
\begin{equation}
E_l=E_o-\delta_1-\Delta_2. 
\end{equation}
Scenario-III: In this scenario the cluster relaxes again to level-C, then, to level-D losing phonons. The energy difference now between level-B and level-D is $\delta_2=E_2 -E_0^1=6.4J-4H_{mf}$ where $E_0^1$ is the lowest eigenstate in level-C to which the cluster relaxes. The cluster de-excites optically from this level to level-F if the band electron combines with the cluster as spin-up and to the level-E if the band electron combines with the cluster as spin-down. The cluster relaxes again losing phonons till it gets the ground state energy level-A. Where the energy difference between level-F and level-A is $\Delta_2$ and the energy difference between level-E and level-A is $\Delta_3=E_{-1/2}^{1,2,3}-E_{5/2}$. Then the optical energy for the scenario-3 is for the first case 
\begin{equation}
E_l=E_0-\delta_2-\Delta_2,
\end{equation}
and for the second case, 
\begin{equation}
E_l=E_0-\delta_2-\Delta_3.
\end{equation}
Because $\Delta_1$, $\Delta_2$ and $\Delta_3$ are too complicated it is not possible to put them here as function of the exchange parameters.
\section{\label{conclusion}Conclusion}
A new physical effect, namely a spin Frank-Condon Effect, has been proposed. A simple model of a spin cluster has been defined that shows rich physics. The energy states of this small spin cluster problem have been investigated and the optical excitation and fluorescence calculated. It was found that in this case the selection rules imposed a very strict limitation on the number of states that could be observed. The physical reason that this occurs is that the lowest energy state for a given spin of the four-spin cluster is always even with respect to permutations of the four sites. Only one state of the five-spin cluster respects this symmetry so only one transition is allowed.

In all cases we found that the cluster decayed to a unique state in ground state even when, as for a ground state of total spin $S_T=1/2$, there are as many as four energies and ten states corresponding to $S_T=\pm1/2$.

Real physical situations will be more complicated. A big simplification here was that the ground state was ferromagnetically aligned and hence its wavefunction was known and it was non-degenerate. More realistic models would be antiferromagnetic. Also this was a model built from $S=1/2$ spins which is again a simplification. An extension to the study of the $e_g$ orbitals of Mn$^{3+}$ LaMnO$_3$ could be done to extend the work of Allen {\it et al}\cite{Allen99}. It would involve states that were rotated\cite{Deisenhofer03} by $2\pi/3$ and would again be complex

\begin{acknowledgements}
This work is funded by the Egyptian High Education Ministry (EHEM).
\end{acknowledgements}


\end{document}